# Influence of Internal Fields on the Electronic Structure in Self-Assembled InAs/GaAs Quantum Dots

**Sharnali Islam, Sasi Sundaresan, and Shaikh Ahmed**

*Department of Electrical and Computer Engineering*
*Southern Illinois University at Carbondale*
*1230 Lincoln Drive, Carbondale, IL 62901, USA.*
*Phone: (618) 453-7630, Fax: (618) 453-7972, E-mail: ahmed@siu.edu*

*Abstract*—
Built-in electrostatic fields in Zincblende quantum dots originate mainly from—(1) the fundamental crystal atomicity and the interfaces between two dissimilar materials, (2) the strain relaxation, and (3) the piezoelectric polarization. In this paper, using the atomistic NEMO 3-D simulator, we study the origin and nature of the internal fields in InAs/GaAs quantum dots with *three* different geometries, namely, box, dome, and pyramid. We then calculate and delineate the impact of the internal fields in the one-particle electronic states in terms of shift in the conduction band energy states, anisotropy and non-degeneracy in the $P$ level, and formation of mixed excited bound states. Models and approaches used in this study are as follow: (1) Valence force field (VFF) with strain-dependent Keating potentials for *atomistic* strain relaxation; (2) 20-band nearest-neighbor $sp^3d^5s^*$ tight-binding model for the calculation of single-particle energy states; and (3) For piezoelectricity, for the first time within the framework of $sp^3d^5s^*$ tight-binding theory, *four* different recently-proposed polarization models (linear and non-linear) have been considered in conjunction with an atomistic 3-D Poisson solver that also takes into account the image charge effects. Specifically, in contrast to recent studies on similar quantum dots, our calculations yield a *non-vanishing* net piezoelectric contribution to the built-in electrostatic field. Demonstrated also is the importance of full three-dimensional (3-D) *atomistic* material representation and the need for using *realistically-extended* substrate and cap layers (systems containing ~2 million atoms) in the numerical modeling of these reduced-dimensional quantum dots.

## I. INTRODUCTION

Rapid progress in nanofabrication technology has made possible the growth of various nanoscale devices where both the atomicity and quantum-mechanical effects play a critical role in determining the overall device characteristics. This leads to a considerable challenge in modeling these devices. The lack of structural symmetry in the overall geometry of the nanodevices usually requires explicit three-dimensional representation. For example, Stranski-Krastanov growth techniques tend to produce self-assembled InAs/GaAs quantum dots (QDs) [1] [2] with some rotational symmetry, *e.g.* disks, domes, or pyramids. These structures are generally not perfect geometric objects, since they are subject to interface inter-diffusion and discretization on an atomic lattice. There is no such thing as a round disk on a crystal lattice! Therefore, the underlying crystal/atomistic asymmetry imposes immediate restrictions on the realistic geometry and demands a full atomistic treatment.

*Strain* originates from the assembly of lattice-mismatched semiconductors and, in the Stranski-Krastanov growth mode, indeed drives the creation of the QDs. In the case of the InAs/GaAs quantum dots, the lattice mismatch is around 7% and leads to a strong *long-range* strain field within the extended neighborhood of each quantum dot [3]. Strain can be atomistically inhomogeneous, involving not only *biaxial* components but also non-negligible *shear* components. Strain in reduced dimensional structures strongly influences the core and barrier material band structures, modifies the energy bandgaps, and further lowers the underlying crystal symmetry. In the nanoscale regime, the classical harmonic linear/continuum elasticity model for strain, which can capture strain only on a mesoscopic scale, is clearly inadequate [4] [5], and device simulations must include the fundamental quantum character of charge carriers and the long-ranged atomistic strain effects with proper boundary conditions on an equal footing.

A variety of materials such as GaAs, InAs, GaN, are *piezoelectric*. Any spatial non-symmetric distortion in nanostructures made of these materials will create piezoelectric fields, which will modify the electrostatic potential landscape. Recent spectroscopic analyses of self-assembled QDs demonstrate polarized transitions between confined hole and electron levels [2]. While the continuum models (effective mass or $k \bullet p$) can reliably predict aspects of the single-particle energy states, they fail to capture the observed non-degeneracy and optical anisotropy of the excited energy states in the (001) plane. These methods fail because they use a confinement potential which is assumed to have only the *structural symmetry* of the nanostructure, and they ignore the underlying crystal asymmetry. The experimentally measured symmetry is significantly lower than the assumed continuum/shape symmetry mainly because of underlying crystalline atomicity and interfaces, strain relaxation, and the piezoelectric fields. For example, in the case of pyramidal QDs with square bases, continuum models treat the underlying material in $C_{4v}$ symmetry while the atomistic representation lowers the crystal symmetry to $C_{2v}$ [2]. QDs with circular bases having structural $C_\infty$ symmetry also exhibit optical polarization anisotropy due to the atomistic asymmetry and the built-in electrostatic fields induced in the underlying lattice.

In this paper, we study the electronic properties of Zincblende InAs quantum dots grown on GaAs substrate. The main objectives are three-fold—(1) to explore the nature and the role of crystal atomicity at the interfaces, strain-field, and piezoelectric polarization in determining the energy spectrum and the wavefunctions, (2) to address shift in the one-particle energy states, symmetry-lowering and non-degeneracy in the first excited state, and strong band-mixing in the overall conduction band electronic states, a group of inter-related phenomena that has been revealed in recent spectroscopic analyses, and (3) to study the geometry-dependence of the above-mentioned phenomena. Efforts are made to demonstrate the importance of three-dimensional (3-D) *atomistic* material representation, and the need for using *realistically-extended* substrate and cap layers in studying the built-in structural and electric fields in these reduced-dimensional QDs. The paper is organized as follows—In Sec. II, we outline the methods for the calculation of strain and piezoelectric fields, and single-particle electronic states. In Sec. III, the internal fields and the electronic structures of InAs/GaAs QDs have been delineated as a function of their geometry. Conclusions are drawn in Sec. IV.

## II. MODELS

It is clear that, at nanoscale, modeling approaches based on a *continuum* representation (such as effective mass [6], and $k \bullet p$ [7]) are clearly invalid. On the other side, various *ab initio* atomistic materials science methods (fundamental many-electron correlated methods based on perturbation theory, quantum Monte Carlo method, or GW approach) offer intellectual appeal, but can only predict masses and bandgaps for very small systems (around 100 atoms). Thus, for quantum dot simulations, the simulation domain requiring multimillion atoms prevent the use of *ab initio* methods. Empirical methods (*Pseudopotentials* [2] [8] and *Tight Binding* [9]), which eliminate enough unnecessary details of core electrons, but are finely tuned to describe the atomistically dependent behavior of valence and conduction electrons, are attractive in realistically-sized nanodevice simulations. Tight-binding is a local basis representation, which naturally deals with finite device sizes, alloy-disorder and hetero-interfaces and it results in very sparse matrices. The requirements of storage and processor communication are therefore minimal compared to pseudopotentials and implementations perform extremely well on inexpensive Linux clusters.

This study has been performed through atomistic simulations using the Nanoelectronic Modeling tool NEMO 3-D. Detail description of this package can be found in Ref. [10]. Based on the atomistic valence-force field (VFF) method [11] for the calculations of strain fields and a variety of tight-binding models ($s$, $sp^3s^*$, $sp^3d^5s^*$) [9] [12] [13] optimized with genetic algorithms to match experimental and theoretical electronic structure data, NEMO 3-D enables the computation of atomistic (non-linear) strain and piezoelectric field for over 64 million atoms and of electronic structure for over 52 million atoms, corresponding to volumes of $(110nm)^3$ and $(101nm)^3$, respectively. Excellent parallel scaling up to 8192 cores has been demonstrated [14]. We are not aware of any other semiconductor device simulation code that can simulate such large number of atoms. NEMO 3-D includes spin in its fundamental atomistic tight binding representation. Effects of interaction with external electromagnetic fields are also included [15]. NEMO 3-D developers are currently making efforts to include structural relaxation in large systems using a hybrid quantum mechanics and molecular mechanics (QM/MM) scheme where part of the system (such as the interface region with atoms close to surfaces) is treated with a full density functional method while the remainder of the system is treated using the tight binding scheme [16]. Recent work on multi-million atom simulations demonstrates the capability of NEMO 3-D to model a large variety of relevant, realistically sized nanoelectronic devices [10] [14] [17] including bulk materials, quantum dots, quantum wires, quantum wells and nanocrystals.

The piezoelectric polarization **P** is obtained from the *shear* components of the stress fields. For the calculations of the piezoelectric polarization, we have considered *four* different models and followed the recipe in Refs. [2] and [18]—(1) within a linear approximation using *experimental* (bulk) values for polarization constants (−0.045 C/m² for InAs, and −0.16 C/m² for GaAs); (2) within a linear/first-order approximation using microscopically-determined (*ab initio* calculations using density functional theory) values for polarization constants (−0.115 C/m² for InAs, and −0.230 C/m² for GaAs); (3) second-order (quadratic) polarization using microscopically-determined values for polarization constants ($\beta_{114} = -0.531$, $\beta_{124} = -4.076$, $\beta_{156} = -0.120$ for InAs, and $\beta_{114} = -0.439$, $\beta_{124} = -3.765$, $\beta_{156} = -0.492$ for GaAs); and (4) A combination of first and second order effects using the above mentioned microscopically-determined values for polarization constants. After calculating the polarization, the piezoelectric *charge* density is derived by taking divergence of the polarization. To do this, we divided the simulation domain into cells by rectangular meshes. Each cell contains four cations. The polarization of each *grid* is computed by taking an average of *atomic* (cations) polarization within each cell. A finite difference approach is then used to calculate the charge density by taking divergence of the grid polarization. Finally, the piezoelectric potential is determined by the solution of the Poisson equation. We have used a PETSc-based [19] *parallel* full 3-D Poisson solver that can capture the image charge effects and is particularly suitable for atomistically resolved Zincblende lattices.

### III. SIMULATION RESULTS

*Figure 1* shows the simulated quantum dots having box, dome and pyramid geometries. The InAs QDs grown in the [001] direction and embedded in a GaAs substrate used in this study have diameter/base length, $d\sim11.3$ nm and height, $h\sim5.6$ nm, and are positioned on an InAs wetting layer of one atomic-layer thickness. The simulation of strain is carried out in the large computational box, while the electronic structure computation is restricted to the smaller inner domain. All the strain simulations fix the atom positions on the bottom plane to the GaAs lattice constant, assume periodic boundary conditions in the lateral dimensions, and open boundary conditions on the top surface. The inner electronic box assumes a closed boundary condition with passivated dangling bonds.

*Figure 2* shows the first four conduction band wavefunctions for each of the quantum dots *without* any strain relaxation and piezoelectricity. Both the InAs dot and the GaAs barrier assume the lattice positions of perfect Zincblende GaAs. In this picture, the geometric shape symmetry is broken and the quantum dots, since constructed from atoms, have $C_{2v}$ symmetries [2]. From the P level wavefunctions, it is clear that the fundamental crystal atomicity and the interfaces (between the dot material InAs and the barrier material GaAs) lower the geometric shape symmetry even in the absence of strain. The interface plane cannot be treated as a reflection plane [2] and creates a short-range interfacial potential. However, it is important to note that the magnitude of the split (non-degeneracy) in the P level is largest in the case of a dome dot ($\sim7.39$ meV) and minimum in a box ($\sim1.2$ meV). Also, the anisotropy in the P level assumes opposite *orientations* for dome and pyramid dots. The first P state is oriented along [1$\underline{1}$0] ([110]) direction and the second along [110] ([1$\bar{1}$0]) direction in a dome (pyramid). For a box, however, the P states are almost isotropic although the degeneracy is somewhat broken (~1.2meV).

Next, we introduce atomistic strain relaxation in our calculations using the VFF method with the Keating Potential. In this approach, the total elastic energy of the sample is computed as a sum of bond-stretching and bond-bending contributions from each atom. The equilibrium atomic positions are found by minimizing the total elastic energy of the system. However, piezoelectricity is neglected in this step. The total elastic energy in the VFF approach has only one global minimum, and its functional form in atomic coordinates is quartic. The conjugate gradient minimization algorithm in this case is well-behaved and stable. *Figure 3a* shows the convergence characteristic in a typical simulation of atomistic strain relaxation of an InAs/GaAs quantum dot with ~1.8 millions of atoms. Strain modifies the effective confinement volume in the device, distorts the atomic bonds in length and angles, and hence modulates the confined states. *Figure 3b* shows the trace of the hydrostatic strain distribution along the [001] direction for all three quantum dots (cut through the center of the dot). There are three salient features in this plot—(1) Atomistic strain is long-ranged and penetrates deep into both the substrate and the cap layers. However, the spread is different in different-shaped quantum dots—largest in a box and minimum in a Pyramid. This may be attributed to the varying total volumetric pressure on to the surrounding material matrix. The spread of strain deep inside the surrounding material matrix not only stresses the importance of including enough of a substrate and cap layer to capture the long-range strain but indicates opportunities to tune the energy spectrum with different capping layer thicknesses also. (2) In all three quantum dots, the hydrostatic strain has non-zero slopes within the quantum dot region. The presence of the gradient in the hydrostatic strain introduces unequal stress in the Zincblende lattice structure along the depth, breaks the equivalence of the [110] and [1$\bar{1}$0] directions, and finally breaks the degeneracy of the P level [2]. *Figure 3c* shows the wavefunction distributions for the first 4 (four) conduction band electronic states in a 2-D projection. Noticeable is the pronounced optical anisotropy and non-degeneracy in the P level. It is also important to note that strain introduces uniform orientational pressure in all three quantum dots. For all three quantum dots, the first P state is oriented along the [110] direction and the second along the [1$\bar{1}$0] direction. *Figure 3c* also reveals the split/non-degeneracy in the P level (defined as $\Delta P = E_{1\bar{1}0} - E_{110}$) in each of the quantum dots. This value is found to be largest in a pyramid and smallest in a box.

In pseudomorphically grown semiconductor heterostructures, the presence of non-zero atomistic stress tensors results in a deformation in the crystal lattice and leads to a piezoelectric field, which has been incorporated in the Hamiltonian as an external potential (within a non-selfconsistent approximation) by solving the Poisson equation on the Zincblende lattice. Traditionally, piezoelectric field in Zincblende quantum dots is thought to be originating from the contribution of the *shear* components of the strain tensors only [2] [14]. *Figure 4* shows the atomistic shear strain profiles in all three quantum dots. The off-diagonal strain tensors have the largest *magnitude* in the dome and minimum in the box shaped dots. However, it is important to note that the shear stress is maximally *spread* in a box dot (~22 nm within the substrate material), which also suggests that piezoelectric contribution would be largest in a box shaped quantum dot. The assumption that piezoelectric fields originate only from the *shear* components of the strain tensors is sometimes referred to as the *linear/first-order* treatment of piezoelectricity, which has recently been revised for InAs/GaAs quantum dots subject to the presence of enormous strain fields [18] [20]. Authors in Ref. [20] have calculated the piezoelectric properties of self-assembled Zincblende quantum dots using both linear and quadratic piezoelectric tensors that are derived from first-principles density functional theory. They have found that the previously ignored second-order term has comparable magnitude as the linear term and the two terms tend to oppose each other. Motivated by this work, in our calculations, we have included 4 (four) different models for piezoelectric polarization as mentioned in the previous section. The resulting piezoelectric potential distributions, corresponding to these 4 (four) different models, along the *Z* (growth) direction are shown in *Figure 5* for all three quantum dots. From this Figure, one can extract at least three important features—(1) Piezoelectric potential has its largest *magnitude* in a pyramidal dot with the peak being located near the pyramid tip, and the minimum in a box. (2) As in the case of strain, the spread of the potential is largest in a box and minimum in a pyramid. (3) *Within the quantum dot region*, the second-order effect has comparable/similar magnitude as the first-order contribution, and, indeed, the two terms oppose each other. However, noticeable is the fact that the first-order contribution, as compared to the quadratic term, penetrates deeper inside the surrounding material matrix. This particular effect, we believe, in contrast to the findings in Ref. [20], results in a non-vanishing and reasonably large *net* (1st+2nd) piezoelectric potential within the region of interest. The fact that the 1st order and the 2nd order terms oppose each other is also noticeable in *Figure 6*, which depicts the surface plots of the piezoelectric potential distribution in the *XY* plane. Note that the 1st order term has got somewhat larger magnitude and spread than the quadratic term. Also, associated with both these two terms, noticeable is the asymmetry and inequivalence (in terms of potential magnitude and distribution) along the [110] and the [1$\bar{1}$0] directions.

*Figure 7* shows the first 4 (four) conduction band wavefunctions for all three quantum dots including *both* the strain and the piezoelectric fields (4th model) in the calculations. The piezoelectric potential introduces a global shift in the energy spectrum and generally opposes the strain induced field. In box and dome shaped dots, the net piezoelectric potential is found to be strong enough to fully offset the combined effects of interface and strain fields and, thereby, *flip* the optical polarization anisotropy. Also shown in *Figure 7* are the splits in the P levels ($\Delta$P) for all three quantum dots. To fully assess the piezoelectric effects, we have prepared Table I that quantifies the *individual* net contributions from crystal atomicity and interfaces, strain, and the various components of piezoelectric fields in the spilt of the P level. The net piezoelectric contribution is found to be largest in a box and minimum in a pyramid, which clearly establishes a one-to-one correspondence between the piezoelectric potential and the volume of the quantum dot under study.

## IV. CONCLUSIONS

Atomistic simulations using the NEMO 3-D tool have been performed to study the origin and nature of the built-in electrostatic fields in Zincblende InAs/GaAs quantum dots with 3 (three) different shapes, namely, box, dome, and pyramid, all having a diameter/base length of 11.3 nm and a height of 5.6 nm. The atomistic strain fields (both hydrostatic and shear) are found to be long-ranged and have the largest spread (~15nm) inside the surrounding material matrix for a box structure. As opposed to the atomic/interfacial symmetry, strain is found to have a general/uniform tendency to orient the electronic wavefunctions along the [110] direction and further lower the symmetry of the system under study. The net contribution from the strain field is found to be largest in the dome shaped quantum dot. Regarding piezoelectricity, for the first time, 4 (four) different models for polarization have been implemented within the atomistic tight-binding description. In consistent with Ref. [20], we find that, *within* the quantum dot region, the contributions from the linear and the quadratic terms have comparable magnitudes yet they tend to cancel each other. The quadratic term, therefore, cannot be neglected and must be taken into account. However, in contrast to Ref. [20], our calculations yield a non-vanishing and reasonably large *net* piezoelectric potential, which can be attributed to the fact that the potential from the linear term, as compared to the quadratic counterpart, penetrates deeper into the surrounding material matrix. This particular observation essentially stresses the need for using *realistically-extended* substrate and cap layers (simulation domains containing ~2 million atoms) in the numerical modeling of these reduced-dimensional quantum dots.

## ACKNOWLEDGEMENTS

This work is supported by the ORAU/ORNL High-Performance Computing Grant 2009. The development of the NEMO 3-D tool involved a large number of individuals at JPL and Purdue, whose work has been cited. Currently, the NEMO 3-D code is being maintained by Gerhard Klimeck at Purdue, discussions with whom is acknowledged. The authors would like to thank the NCCS at Oak Ridge National Laboratories, the Rosen Center for Advanced Computing (RCAC) at Purdue University, and the nanoHUB.org for providing computational resources and support.

# LIST OF FIGURES

**Figure 1**. Simulated InAs/GaAs quantum dots on a thin (one atomic layer) InAs wetting layer. Two major computational domains are also shown. $D_{elec}$: central smaller domain for electronic structure (quantum) calculation, and $D_{strain}$: outer domain for strain calculation. In the figure: $s$ is the substrate height, $c$ is the cap layer thickness, $h$ is the dot height, and $d$ is the dot diameter/base length as appropriate.

**Figure 2**. First four conduction band wavefunctions due to fundamental crystal and interfacial symmetry. Noticeable are the split (non-degeneracy) and the anisotropy (dominant in dome and pyramidal dots) in the P level.

**Figure 3**. (a) Convergence of elastic energy in a typical simulation of InAs/GaAs quantum dot with ~2 million atoms. The total elastic energy of the sample is minimized using the conjugate gradient method. (b) Trace of atomistic hydrostatic strain along the growth ([001]) direction through the center in all three quantum dots. Diameter/base length, $d = 11.3$ nm, height, $h = 5.65$ nm, substrate thickness, $s = 30$ nm, and cap layer thickness, $c = 10$ nm. Strain is seen to penetrate deep into the substrate and the cap layers. Also, noticeable is the gradient of strain inside the dot region that results in optical polarization anisotropy and non-degeneracy in the conduction band $P$ level. (c) First 4 (four) electronic wavefunctions and split in the P levels in all three quantum dots including strain relaxation. Number of atoms simulated: 1.78 millions (strain domain), 0.8 million (electronic domain).

**Figure 4**. Atomistic shear strain profiles along the $z$ (growth) direction that, in effect, induces polarization in the sample. Note the varying spread/penetration in the surrounding material matrix as a function of dot shape.

**Figure 5**. Induced piezoelectric potential along the $z$ (growth) direction in all three quantum dots. Four different models for the polarization constants have been used in the calculations [18]: (1) linear and experimentally measured, (2) linear through *ab initio* calculations, (3) quadratic through *ab initio* calculations, and (4) combination of 1st and 2nd order components. Also in this Figure, note the varying spread/penetration of the potential in the surrounding material matrix as a function of dot shape.

**Figure 6**. Linear and quadratic contributions of the piezoelectric potential in the $XY$ plane halfway through the dot height. Note the magnitude, orientation, and anisotropy in the induced potential.

**Figure 7**. First 4 (four) electronic wavefunctions and split in the P levels in all three quantum dots including atomicity/interfacial effects, strain, and piezoelectricity. Note the varying piezoelectric contributions as a function of shape, which can be associated mainly to the volume of the quantum dot under study.

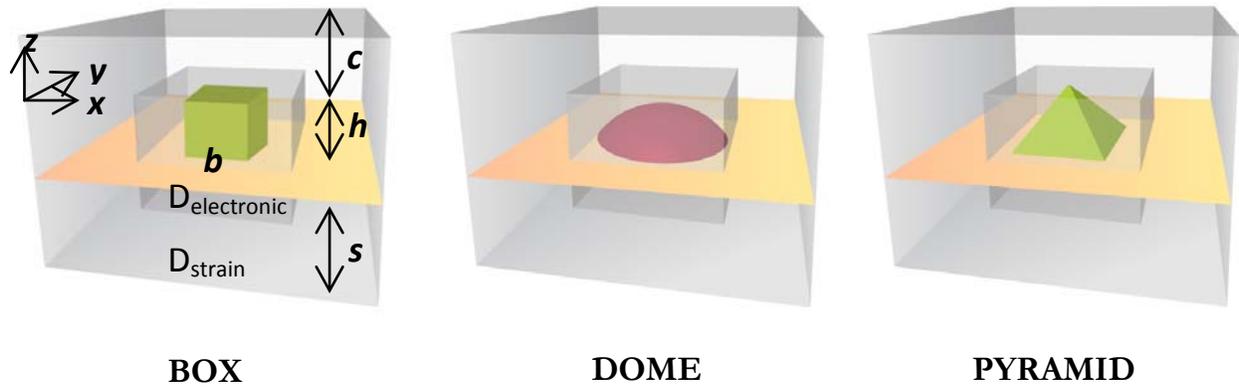

**BOX**          **DOME**          **PYRAMID**

**Figure 1**. Simulated InAs/GaAs quantum dots on a thin (one atomic layer) InAs wetting layer. Two major computational domains are also shown. $D_{elec}$: central smaller domain for electronic structure (quantum) calculation, and $D_{strain}$: outer domain for strain calculation. In the figure: $s$ is the substrate height, $c$ is the cap layer thickness, $h$ is the dot height, and $d$ is the dot diameter/base length as appropriate.

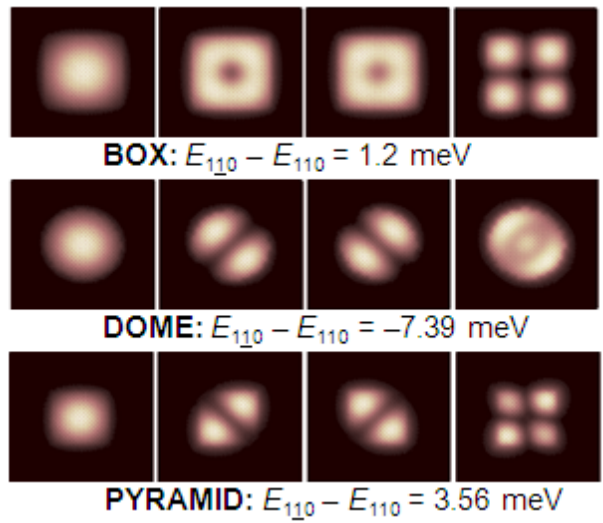

**Figure 2**. First four conduction band wavefunctions due to fundamental crystal and interfacial symmetry. Noticeable are the split (non-degeneracy) and the anisotropy (dominant in dome and pyramidal dots) in the P level.

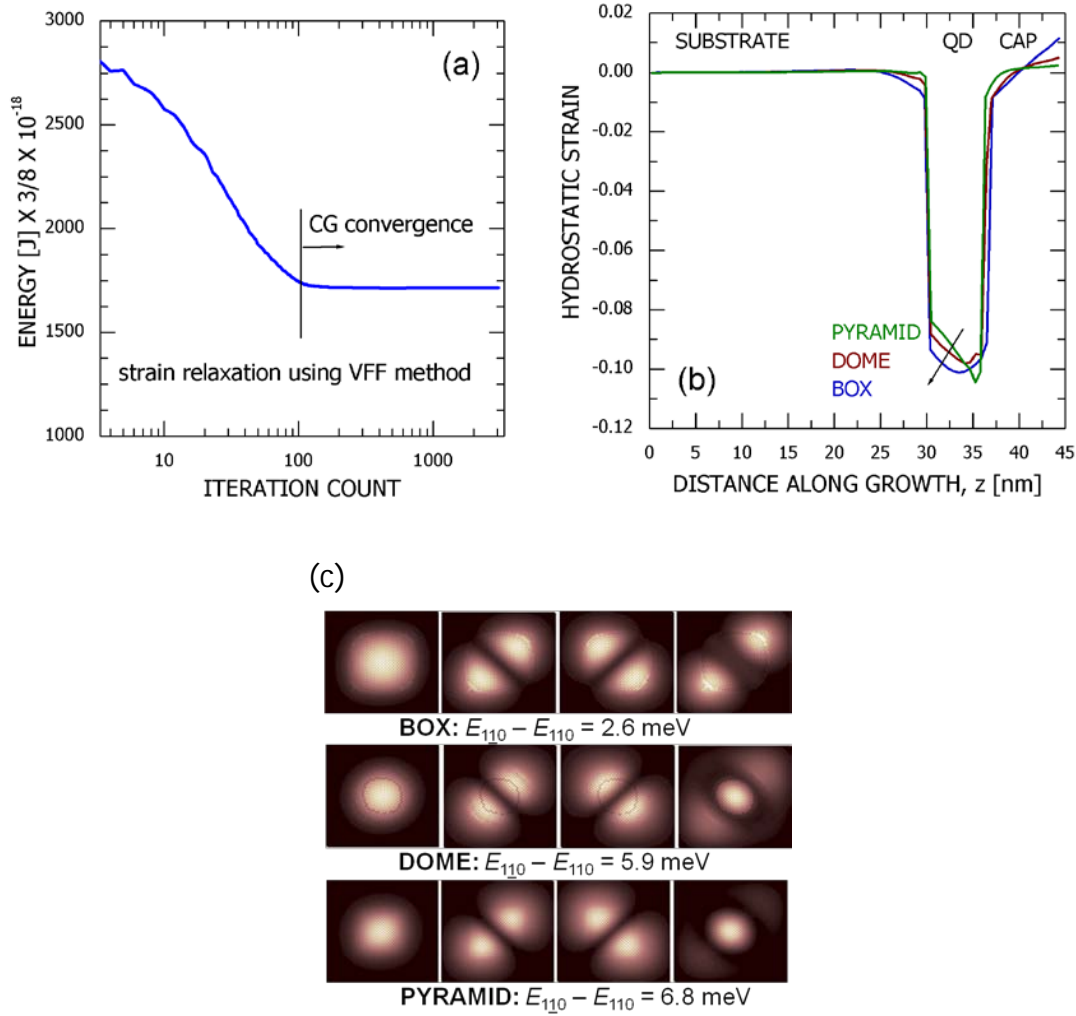

**Figure 3**. (a) Convergence of elastic energy in a typical simulation of InAs/GaAs quantum dot with ~2 million atoms. The total elastic energy of the sample is minimized using the conjugate gradient method. (b) Trace of atomistic hydrostatic strain along the growth ([001]) direction through the center in all three quantum dots. Diameter/base length, $d = 11.3$ nm, height, $h = 5.65$ nm, substrate thickness, $s = 30$ nm, and cap layer thickness, $c = 10$ nm. Strain is seen to penetrate deep into the substrate and the cap layers. Also, noticeable is the gradient of strain inside the dot region that results in optical polarization anisotropy and non-degeneracy in the conduction band $P$ level. (c) First 4 (four) electronic wavefunctions and split in the P levels in all three quantum dots including strain relaxation. Number of atoms simulated: 1.78 millions (strain domain), 0.8 million (electronic domain).

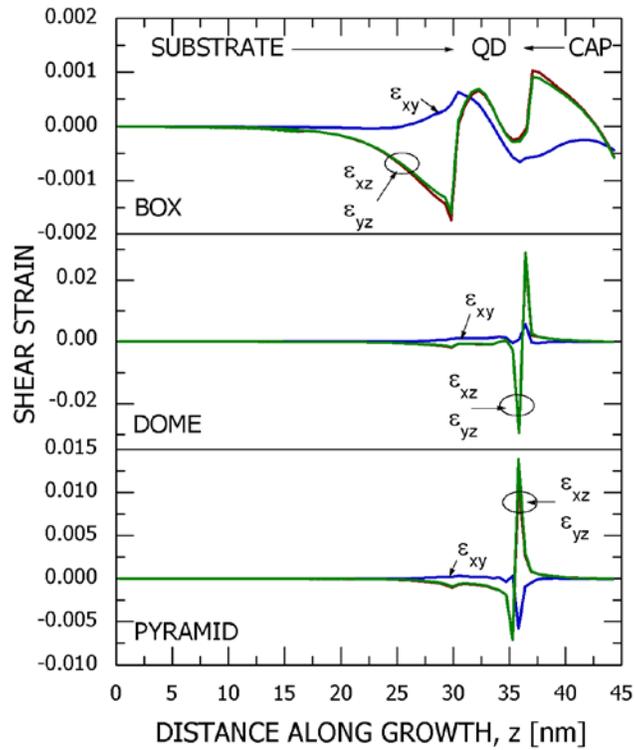

**Figure 4**. Atomistic shear strain profiles along the *z* (growth) direction that, in effect, induces polarization in the sample. Note the varying spread/penetration in the surrounding material matrix as a function of dot shape.

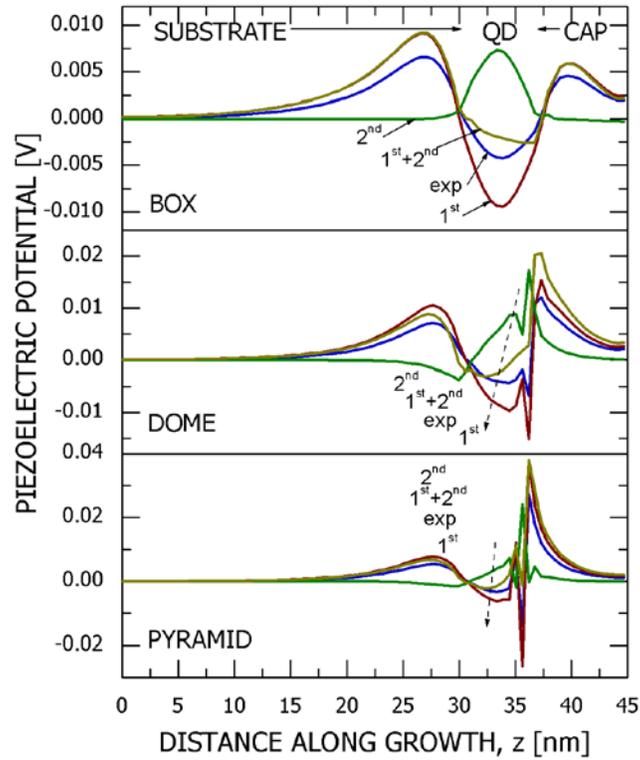

**Figure 5**. Induced piezoelectric potential along the $z$ (growth) direction in all three quantum dots. Four different models for the polarization constants have been used in the calculations [18]: (1) linear and experimentally measured, (2) linear through *ab initio* calculations, (3) quadratic through *ab initio* calculations, and (4) combination of 1st and 2nd order components. Also in this Figure, note the varying spread/penetration of the potential in the surrounding material matrix as a function of dot shape.

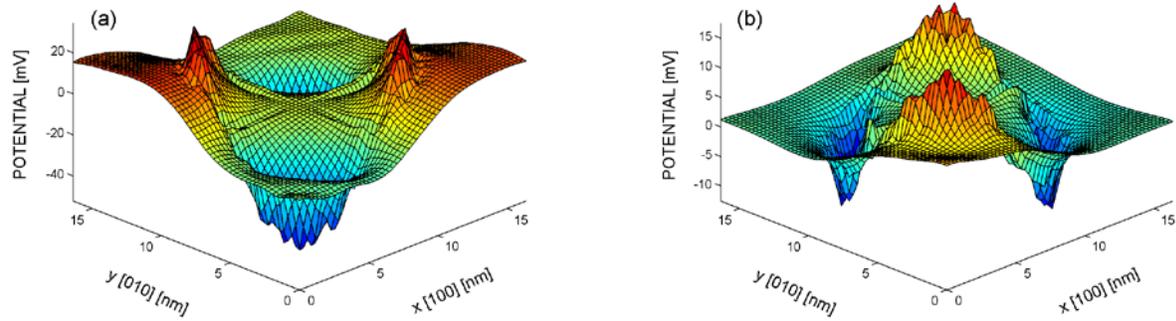

**Figure 6**. Linear and quadratic contributions of the piezoelectric potential in the *XY* plane halfway through the dot height. Note the magnitude, orientation, and anisotropy in the induced potential.

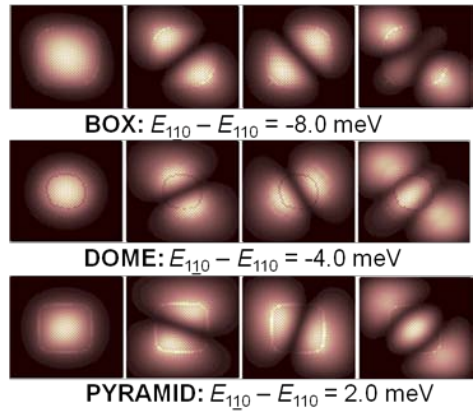

**Figure 7**. First 4 (four) electronic wavefunctions and split in the P levels in all three quantum dots including atomicity/interfacial effects, strain, and piezoelectricity. Note the varying piezoelectric contributions, which can be attributed mainly to the volume of the quantum dot under study

TABLE I: NET CONTRIBUTION (in meV) OF VARIOUS EFFECTS IN SPLITTING THE P LEVEL.

| Effect | BOX | DOME | PYRAMID |
|---|---|---|---|
| Atomicity/Interface | 1.2 | -7.4 | 3.56 |
| Strain relaxation | 1.4 | 13.3 | 3.24 |
| Interface + Strain | 2.6 | 5.9 | 6.8 |
| PZ(1st order) | -12.6 | -16.9 | -10.8 |
| PZ(2nd order) | 0.4 | 5.1 | 5.2 |
| PZ(1st + 2nd order) | -10.6 | -9.9 | -4.8 |